\documentclass{mem}
\usepackage{natbib}\usepackage{txfonts}\usepackage{balance}
\usepackage{graphicx}
\usepackage{hyperref}
\idline{75}{282}
\begin{document}
\def\teff{$T\rm_{eff }$}
\def\kms{$\mathrm {km s}^{-1}$}

\title{
Physical parameter determinations of young Ms}

   \subtitle{Taking advantage of the Virtual Observatory to compare methodologies}

\author{
A. Bayo\inst{1, 2} 
\and C. Rodrigo\inst{3, 4}
\and D. Barrado\inst{3}
\and F. Allard\inst{5}
          }

\institute{
Max-Planck-Institut f\"ur Astronomie, Heidelberg, Germany
\and
Departamento de F\'isica y Astronom\'ia, Facultad de Ciencias, Univ. de Valpara\'iso, Chile
\and
Centro de Astrobiolog\'ia, INTA-CSIC, Madrid, Spain
\and
Spanish Virtual Observatory, Spain
\and
Centre de Recherche Astronomique de Lyon (CRAL), Ecole Normale Superieure de Lyon, France
\email{bayo@mpia.de}
}

\authorrunning{Bayo }

\titlerunning{Young Ms in the VO}

\abstract{
One of the very first steps astronomers working in stellar physics perform to advance in their studies, is to determine the most common/relevant physical parameters of the objects of study (effective temperature, bolometric luminosity, surface gravity, etc.). Different methodologies exist depending on the nature of the data, intrinsic properties of the objects, etc. One common approach is to compare the observational data with theoretical models passed through some simulator that will leave in the synthetic data the same imprint than the observational data carries, and see what set of parameters reproduce the observations best. Even in this case, depending on the kind of data the astronomer has, the methodology changes slightly. After parameters are published, the community tend to quote, praise and criticize them, sometimes paying little attention on whether the possible discrepancies come from the theoretical models, the data themselves or just the methodology used in the analysis. In this work we perform the simple, yet interesting, exercise of comparing the effective temperatures obtained via SED and more detailed spectral fittings (to the same grid of models), of a sample of well known and characterized young M-type objects members to different star forming regions and show how differences in temperature of up to 350 K can be expected just from the difference in methodology/data used. On the other hand we show how these differences are smaller for colder objects even when the complexity of the fit increases like for example introducing differential extinction. To perform this exercise we benefit greatly from the framework offered by the Virtual Observatory.

%
\keywords{Stars: low-mass, brown dwarfs Ð Stars: fundamental parameters Ð Infrared: stars Ð Stars: formation Ð Astronomical databases Ð Virtual observatory tools }
}
\maketitle{}

\section{Introduction}

The determination of stellar  and substellar physical parameters is a mandatory step in order to attack most of the main open questions in star formation: the universality of the Initial Mass Function, the evolution of circumstellar disks, clusters and associations, etc. In addition, almost the totality of the inferred properties on such a hot topic as is exoplanets, relies strongly on the determination of the parameters of their host stars. 

The long tradition of studies of the sun and solar analogs, translates in a ``better" (more detailed) understanding of the physical processes involved in such objects and therefore, models of atmospheres can reproduce to great detail the observed features for such stars. 

On the other hand, very low-mass stars and substellar objects are not so well understood yet, and in particular their cooler atmospheres pose a challenge regarding chemistry (see for example \citealt{Allard12} and references therein) where as dust settles in the atmospheres, different species condensate at different temperatures, etc.  Large advances have been made in the last ten years since the atmosphere and interior model grids with limiting dust settling treatments were proposed (see \citealt{Chabrier00}, \citealt{Allard01}, \citealt{Tsuji02} and \citealt{Baraffe03} as examples) to try to include in the atmospheric codes these complicated cloud models.

A positive feature of very low mass-stars and brown dwarfs is that they do not belong anymore to a class of ``rare" objects, offering large samples to test the models. However, this wealth of data (models in development and more and more observations) can become complicated to analyze efficiently if one does not posses the right tools.

In this context, the Virtual Observatory (VO), based in establishing standards for the publication and exchange of astronomical data, and all the recently developed VO-Tools, can be seen as the necessary bridge to access and understand those data in an efficient manner.

In this work we take advantage of the VO framework to derive the effective temperature of a sample of known members to three star forming regions and young associations. We obtain these temperatures using different sets of data, photometry and spectroscopy, and a single collection of models and compare the difference arising from the kind of data used. The interesting aspect of using these VO-Tools is that they can be used directly on any other collection of models available in the VO in the same manner to learn more about the advantages and disadvantages of each one of these collections (see for example Bayo et al. 2014, submitted where we compare the results of the new grids from the Lyon group with those assuming limiting cases in the dust settling).

\section{Sample and data}

For this exercise we have selected a subsample of the M-class population of Collinder 69 (the central cluster of the Lambda Orionis star forming region, 5-10 Myrs old, \citealt{Barrado04}) spectroscopically confirmed as members to the cluster in \cite{Bayo11}. To this sample, in order to further populate the latest spectral types, we have added late-M members to the Chamaleon I dark cloud, Cha I from \cite{Comeron00} (age $\sim$2 Myrs, citealt{Luhman07}) and the TW-Hydra association (TWA, estimated ages between 8 and 20 Myrs see for example \citealt{Soderblom98, Gizis02, Barrado06}, and references therein).

\begin{table}
\caption{Main properties of the sample of young very low-mass stars and
  brown dwarfs in this work that belong to C69, Cha I or TWA }
\begin{scriptsize}
\begin{center}
\begin{tabular}{@{\extracolsep{-7pt}}lllllrr}
ID & T$_{\rm eff}$$^{a)}$ & Max$\lambda$$^{b)}$ & SpT$^{c)}$ & T$_{\rm eff}$$^{d)}$ & Disk$^{e)}$ & Acc.$^{e)}$\\
   & (spec)  	   & ($\mu$m)  & 	& (SED)	  	& type &\\
\hline
\hline
IRAC-003    & 3100 &   1.03 & M4      & 3500 & Thin       &   N\\ 
IRAC-001    & 3400 &   1.03 & M1.5    & 3600 & Thick      &   N\\ 
IRAC-002    & 3200 &   1.03 & M3      & 3900 & Thick      &   N\\ 
IRAC-006    & 3400 &   0.72 & M3.5    & 3600 & Thick      &   N\\ 
IRAC-007    & 3400 &   0.72 & M2.5    & 3700 & Thin       &   Y\\ 
DM016           & 3700 &   0.72 & M1.5    & 3750 & None   &   N\\ 
LOri038         & 3400 &   0.72 & M3      & 3500 & Thick      &   Y\\ 
LOri043         & 3400 &   0.72 & M4      & 3500 & Trans. &   N\\ 
LOri053         & 3200 &   0.72 & M5      & 3500 & None   &   N\\ 
LOri054         & 3200 &   1.03 & M5.5    & 3700 & None   &   N\\ 
LOri073         & 3000 &   0.72 & M5.25   & 3700 & None   &   N\\ 
LOri079         & 3000 &   0.72 & M6.25   & 3600 & None   &   N\\ 
LOri082         & 3400 &   1.03 & M4.75   & 3500 & None   &   N\\ 
LOri087         & 3200 &   1.03 & M5      & 3500 & None   &   N\\ 
LOri095         & 2800 &   1.03 & M6      & 3400 & None   &   N\\ 
LOri099         & 3000 &   0.72 & M5.5    & 3400 & None   &   N\\ 
LOri109         & 2600 &   0.72 & M5.75   & 3300 & None   &   N\\ 
LOri126         & 2700 &   1.03 & M6.5    & 3100 & Thick      &   Y\\ 
LOri130         & 3200 &   1.03 & M5.25   & 3200 & None   &   N\\ 
LOri139         & 2800 &   1.03 & M5.75   & 3100 & Thick      &   N\\ 
LOri140         & 2700 &   1.03 & M7.0    & 2900 & Thick      &   Y\\ 
LOri155         & 2600 &   1.03 & M8.0    & 2800 & Thin       &   N\\ 
Cha H$\alpha$ 1 & 2500 &   2.35 & M7.75   & 2600 & Thick      &   N\\ 
Cha H$\alpha$ 2 & 2850 &   2.35 & M5.25   & 3100 & Thick      &   N\\ 
Cha H$\alpha$ 3 & 2650 &   2.35 & M5.5    & 2700 & Trans. &   N\\ 
Cha H$\alpha$ 4 & 2800 &   2.35 & M5.5    & 2900 & Trans. &   N\\ 
Cha H$\alpha$ 5 & 2800 &   2.35 & M5.5    & 2900 & None   &   N\\ 
Cha H$\alpha$ 6 & 2500 &   2.35 & M5.75   & 2800 & Thick      &   Y\\ 
Cha H$\alpha$ 7 & 2150 &   2.35 & M7.75   & 2400 & None   &   N\\ 
Cha H$\alpha$ 8 & 2600 &   2.35 & M5.75   & 3000 & Trans. &   N\\ 
Cha H$\alpha$ 9 & 2800 &   2.35 & M5.5    & 3000 & Thick      &   N\\ 
Cha H$\alpha$ 11& 2450 &   2.35 & M7.25   & 2600 & None   &   N\\ 
Cha H$\alpha$ 12& 2100 &   2.35 & M6.5    & 2500 & Thin       &   N\\ 
SSSPMJ1102      & 2500 &   2.35 & M8.5    & 2400 & Thin       &   N\\ 
2MJ1207         & 2500 &   2.35 & M8      & 2500 & Thin       &   N\\ 
2MJ1139         & 2450 &   2.35 & M8      & 2400 & None   &   N\\ 
\hline
\end{tabular}
\end{center}
\vspace{-0.1cm}
\noindent$^{\mathrm{a)}}$ From this work or Bayo et al. (2014, submitted).\\
$^{\mathrm{b)}}$ Maximum wavelength of the avilable spectra.\\
$^{\mathrm{c)}}$ From \cite{Bayo11a} for C69, \cite{Luhman08} for Cha I, and \cite{Gizis02} for TWA \\
$^{\mathrm{d)}}$ From \cite{Bayo11a} or Bayo et al. (2014, submitted).\\
$^{\mathrm{e)}}$ From \cite{Bayo12} or Bayo et al. (2014, submitted).
\end{scriptsize}
\end{table}

The main properties of the targets (and the results from the fits) are summarized in Table 1 and the distribution of their spectral types displayed in Fig. 1. From Cha I and TWA we selected a total of 14 sources for which we have near-infrared (NIR) and or optical low resolution spectroscopy as well as multi-wavelength photometry covering from the optical to the far infrared (FIR). The photometry and spectroscopy of this sample is described in detail in Bayo et al. (2014, submitted) but in short consist of broadband optical, NIR and MIR photometry and low resolution spectroscopy from \cite{Comeron00} and NTT/SOFI NIR low-resolution NIR spectroscopy from \cite{Bayomasters}.

In the case of Collinder 69 (C69), where we had optical spectroscopy for a larger sample of sources from \cite{Bayo11}, we have selected those in the M-class, not reported as tight binaries, with reliable flux calibration, no veiling (from \citealt{Bayo12}) and a wavelength coverage large enough so that a meaningful fit to the continuum can be made in order to estimate the effective temperature ($\sim$1000 \AA). The sample contains disk and diskless sources, and in the case of the former accreting and not accreting objects.

The main difference (with an impact in the physical parameter estimation) among the three regions is that they cover from heavily heterogeneously extinct environments as Cha I to extinction free, close-by objects from TWA passing the lightly homogeneously extinct cluster C69 \citep{Duerr82}, adding a different level of complexity/degeneration to the model fitting process when introducing the visual extinction (A$_{\rm V}$) and/or extinction-law itself as free parameters.


\begin{figure}[t!]
\resizebox{\hsize}{!}{\includegraphics[clip=true]{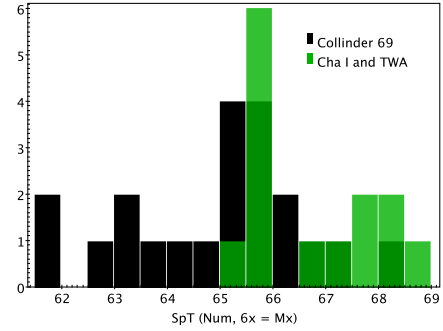}}
\caption{\footnotesize
Spectral type distribution of the members of Collinder 69 selected for the comparisons in this work. The spectral types have been coded numerically where the number six corresponds to the spectral class ``M".
}
\label{LF}
\end{figure}


\section{Results}

\begin{figure}[t!]
\resizebox{\hsize}{!}{\includegraphics[clip=true]{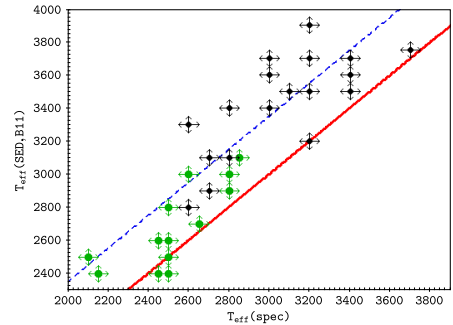}}
\caption{\footnotesize
Comparison of T$_{\rm eff}$ determined using the full SED and those derived fitting the optical spectrum (see text for details). Typical 100 K step in the model grids is illustrated with double pointed arrows. Green solid circles for the Cha I and TWA objects (estimations from Bayo et al 2014, see text) and black circles for the C69 sample (SED estimated temperatures from \citealt{Bayo11a} and spectral fittings from this work). The red solid line marks the one-to-one correlation, and the blue dashed line a parallel one but with an added 350 K step.
}
\label{comparetef}
\end{figure}

By using VOSA\footnote{\scriptsize\url{http://svo2.cab.inta-csic.es/svo/theory/vosa/}} \citep{Bayo08}, the SVO synthetic photometry generator/server\footnote{\scriptsize\url{http://svo2.cab.inta-csic.es/svo/theory/myspec/}} and the theoretical spectra one\footnote{\scriptsize\url{http://svo2.cab.inta-csic.es/svo/theory/newov/}},  we performed simple brute-force fitting agains the BT-Settl grid of models \citep{Allard12} to the two sets of data: on the one hand the multi-wavelength Spectral Energy Distributions (SEDs), and on the other hand the optical and/or NIR spectra (where we adapted the synthetic spectra to the corresponding resolution and masked ``problematic" areas of the observed spectra as emission lines coming from activity/accretion and telluric bands). The comparison of the results obtained using the two data-sets are described in the following.


As a first step we compared the results obtained for C69, where the extinction is known a priori and we had optical medium to low resolution spectroscopy, and TWA, where the extinction can be considered negligible and for these objects (three), we only had NIR spectroscopy. This comparison is shown in Fig.2, where we see how the temperatures estimated via SED fit are higher than those determined with the spectra. It is interesting to note that this stands for those samples of objects even though one set of ``spectroscopic temperatures" are determined with optical data, and the other one with NIR. It is also quite remarkable that these differences are in average of the order of 350 K, much larger than the typical 100 K step in temperature provided in the model grids.

When we included the visual extinction as a free parameter (the case of Cha I), but also larger wavelength spectral coverage (optical + NIR spectra), we see that the agreement between both methodologies is much better (2300-2800 K range in Fig.2).


\section{Implications}

To conclude the exercise, we have constructed the HR diagram for the C69 sample (for the sake of clarity we are including the comparison only for one of the samples, Fig. 3) using the two sets of estimations of effective temperatures (and bolometric luminosities) and it is obvious that the masses and ages obtained with each set of parameters will differ greatly depending on the set used for the comparison with isochrones.

On the other hand, features like the luminosity spread, remain regardless of the data-set used to estimate the bolometric luminosity and effective temperature.


\begin{figure}[t!]
\resizebox{\hsize}{!}{\includegraphics[clip=true]{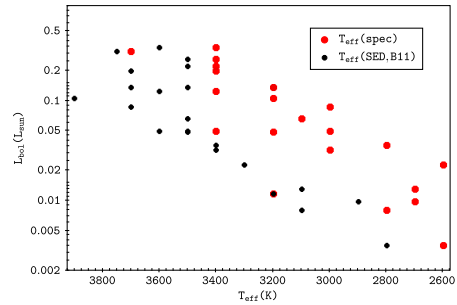}}
\caption{\footnotesize
Comparison of the effective temperatures determined using the full SED and that derived when fitting the optical spectrum masking the areas corresponding to physics not included in the models (i.e. chromosphere). For sake of clarity of the figure we display the situation only for the C69 sample.
}
\label{HR}
\end{figure}

\begin{acknowledgements}
\begin{small}
This publication makes use of VOSA, developed under the Spanish Virtual Observatory project supported from the Spanish MICINN through grant AYA2011-24052.
\end{small}
\end{acknowledgements}
\scriptsize
\bibliographystyle{aa}

\end{document}